\documentclass[12pt]{article}
\begin{document}

\begin{flushright}
{\bf FIAN/TD-10/04}\\{\bf ITEP-TH-36/04}
\end{flushright}

\bigskip

\begin{center}
{\bf Color Glass Condensate in High Energy QCD}
\footnote{Talk given at "QUARKS 2004", May 2004, Pushkinskie Gory, Russia}
\end{center}

\bigskip

\begin{center}
{\bf A.V.~Leonidov$^{(a,b)}$\footnote{Based on work done in collaboration with E.~Ferreiro, J.~Jalilian-Marian,
E.~Iancu, A.~Kovner, L.~McLerran and H.~Weigert.}}
\end{center}
\begin{center}
(a) {\it Theoretical Physics Department, P.N.~Lebedev Physics Institute, Moscow, Russia}

(b) {\it Institute of Theoretical and Experimental Physics, Moscow, Russia}
\end{center}

\bigskip

\begin{center}
{\bf Abstract }
\end{center}

A brief review of major theoretical aspects of color glass condensate physics is given.

\newpage

This talk presents a brief review of some theoretical aspects related to the physics of color glass condensate
(for detailed reviews see, e.g., \cite{ILM02,IV03} with the main focus on the results obtained in
\cite{JKLW97,JKLW99a,ILM00,ILM01,FILM02}. In its essence the color glass condensate physics is based on the
quasiclassical perturbative approach to high energy QCD, in which contributions of all orders of $\alpha_s \ln
1/x$, where $x$ is a fraction of the hadron (nuclear) longitudinal momentum carried by the considered gluon mode,
and of all orders in gluon density are resummed.

The main issue of the perturbative approach to high energy QCD is that of uitarity. Resummation of leading energy
logarithms $O(\alpha_s \ln 1/x)$ in the linear approximation in gluon density through the BFKL equation for the
(unintegrated) gluon density $\phi(Q^2_{\perp},x)$ \cite{BFKL}:
\begin{equation}\label{BFKL}
{\partial \phi \over \partial \ln 1/x} = \alpha_s \, {\cal K}_{BFKL} \cdot \phi \,\,\, \Longrightarrow \phi  |_{x
\to 0} \sim \left ( {1 \over x} \right )^\omega_{BFKL}.
\end{equation}
The growth of the gluon density  and the resulting growth of physical cross-sections with energy ($1/x \sim
\sqrt{s}$) is powerlike - in clear violation of the Froissart bound $\sigma \sim 1/m_{\pi}^2 \ln^2 1/x$ following
from the requirement of unitarity of the theory.

Let us note, that besides color glass condensate approach there exist two other major lines of research addressing
the question of unitarization of high energy perturbative asymptotics of QCD, in particular:
\begin{itemize}
\item{An effective theory including, in addition to quarks and gluons, new degrees of freedom - reggeons \cite{Li95,Li97}.}
\item{Generalization of operator product expansion formalism \cite{B96}.}
\end{itemize}

The notion of color glass condensate has been suggested with the following two analogies in mind:
\begin{itemize}
\item{The small -$x$ gluon modes are characterized by large occupation numbers. Therefore one is tempted to
describe physics of gluons in terms of classical gluon fields, corresponding to the condensation of gluonic
modes.}
\item{Averaging over color charge density is similar to averaging over disorder in random magnetics - spin glasses.}
\end{itemize}

\begin{center}
{\bf Tree level: McLerran-Venugopalan model}
\end{center}

Foundations of color glass condensate physics lie in the quasiclassical tree-level description of the wave
function of large nuclei developed by L.~Mclerran and R.~Venugopalan \cite{MV94}. The gluonic content of this
wavefunction is described in terms of fast (having large $k^+ \sim P^+$, where $P^+$ is a longitudinal momentum of
the nucleus) static sources characterized by the color charge density $\rho^a(x^-,x_\perp)$ and slow (small $k^+
\ll P^+$)
\begin{equation} \label{cleq0}
[D_{\nu}, F^{\nu \mu}]\, =\, \delta^{\mu +} \rho_a(x^-,{\bf x}_\perp)\,.
\end{equation}
The solution of (\ref{cleq0}) reads: ${\cal A}^+ = {\cal A}^- =0$ and
\begin{equation}\label{tpg}
{\cal A}^i(x^-,x_{\perp})\,=\,{i \over g}\, U(x^-,x_{\perp})\,\partial^i U^{\dagger}(x^-,x_{\perp})\,,
\end{equation}
$$
\left\{
\begin{array}{rcl}\label{cs}
U^\dagger (x^-,x_\perp) & = & P {\rm exp} \left \{ ig \int_{x^-_0}^{x^-} dz^- \, \alpha(z^-,x_\perp) \right \}
\\
-\nabla_\perp^2 \alpha & = & U^\dagger \rho U \\
\end{array}
\right.
$$
In what follows we shall also need the following approximate expression for $U^\dagger$ that often enters the
expressions arising in calculations of quantum corrections:
\begin{equation}\label{UTAF}
U^{\dagger}(x^-,x_{\perp})\,\equiv\,
 {\rm P} \exp
 \left \{
ig \int_{-\infty}^{x^-} dz^-\,\alpha (z^-,x_{\perp})
 \right \}\approx \,\theta(x^-)\,\Omega^\dagger(x_{\perp}) + \theta(-x^-),
\end{equation}
Calculation of physical observables is then done through computing the tree level correlators on the classical
solution ${\cal A}^i(x^-,x_{\perp})$:
\begin{equation}\label{clascorr}
 \langle A^i_a(x^+,\vec x)A^j_b(x^+,\vec y) \cdots\rangle_\Lambda\,=\,
 \int {\cal D}\rho\,\,W_\Lambda[\rho]\,{\cal A}_a^i({\vec x}) {\cal A}_b^j({\vec y})\cdots\,
\end{equation}
where $W_\Lambda[\rho]$ is a weight functional for the ensemble of sources at the (longitudinal) scale $\Lambda$.

The main result of the tree-level approach is that already in this approximation the infrared growth of gluon
distributions is tamed by nonlinear corrections \cite{JKLW97,KM98}. This happens at the new scale of the theory -
the saturation scale $Q_s$. Let us illustrate the saturation phenomenon by writing down the asymptotical
expressions for the gluon structure function $G(x,Q^2)=\int^{Q^2} dk_\perp^2 \phi(x,k_\perp^2)/k_\perp^2$:
$$
\left\{
\begin{array}{rcl}
 xG(x,Q^2) (k^2_\perp \gg Q^2_s) & \simeq & N_c \, {\alpha_s C_F \over \pi} \, \ln {Q^2 \over \Lambda^2_{QCD}}\\
 xG(x,Q^2) (k^2_\perp \ll Q^2_s) & \simeq & {1 \over 4 \pi^3 \alpha_s} {N^2_c-1 \over N_c} \pi R^2
 \ln {Q^2_s \over Q^2}\\
\end{array}
\right.
$$

\bigskip

\begin{center}
{\bf Quantum corrections: renormalization group in $\tau=\ln 1/x$}
\end{center}

The starting point for constructing a theory allowing to compute quantum corrections to McLerran-Venugopalan model
is writing down an effective action for high energy QCD in terms of classical sources $\rho$ and gluon fields
$A^\mu$. Such an action was first written in \cite{JKLW97}:
\begin{equation}\label{ACTIONR}
 S[A,\rho]\,=\,- \int d^4x \,{1 \over 4} F_{\mu\nu}^a F^{\mu\nu}_a \,+\,{i \over {gN_c}}
 \int d^3 \vec x\, {\rm Tr}\,\Bigl\{ \rho(\vec x) \,W_{\infty,-\infty}(\vec x)\Bigr\}\,\equiv\,S_{YM}\,+\,S_W.\,\,
\end{equation}
where
\begin{equation}\label{WLINER}
 W_{\infty,-\infty}[A^-](\vec x)\, =\,{\rm T}\, \exp\left[\, ig\int dx^+ A^-(x) \right]
\end{equation}
Note that the effective action (\ref{ACTIONR}) includes the standard Yang-Mills piece $S_{YM}$, as well as a gauge
invariant generalization of the abelian eikonal vertex  $\int d^4x \,\rho_a A^-_a$. As shown in \cite{JKLW97},
this action allows to reproduce the BFKL equation \cite{BFKL} in the small-density limit.

The considered effective theory contains two averagins: over the configurations of the classical source $\rho$ at
the scale $\Lambda$ with the weight functional $W_\Lambda [\rho]$ and usual quantum averaging over the quantum
fluctuations over the gluon fields. For example, the two-point correlator reads
\begin{equation}\label{2point}
 << {\rm T}\,A^\mu(x)A^\nu(y) >> \,\,=\,\int {\cal D}\rho\,\,W_\Lambda[\rho]
 \left\{\frac{\int^\Lambda {\cal D}A \,\,A^\mu(x)A^\nu(y)\,\,
 {\rm e}^{\,iS[A,\,\rho]}} {\int^\Lambda {\cal D}A\,\,{\rm e}^{\,iS[A,\,\rho]}} \right\},
\end{equation}

As shown in \cite{JKLW99a,ILM01,FILM02}, the quantum evolution of gluon correlators is compactly parametrized by
the evolution of the weight functional $W_\Lambda[\alpha]$ (it turns out convenient to make a change of variables
from $\rho$ to $\alpha$ related through the solution of the above-described classical equation of motion):
 \begin{equation}\label{RGEa}
  {\partial W_\tau[\alpha] \over {\partial \tau}}\,=\,
  \alpha_s \left\{ {1 \over 2}
  {\delta^2 \over {\delta \alpha_\tau(x) \delta \alpha_\tau(y)}}
  [W_\tau\nu_{xy}] - {\delta \over {\delta \alpha_\tau(x)}}
  [W_\tau\nu_{x}] \right\}\,.
 \end{equation}
The evolution (\ref{RGEa}) includes new virtual and real kernels $\nu$ and $\nu$. Explicit computations
\cite{ILM01,FILM02} lead to the following simple expressions for them:
 \begin{eqnarray}
  \nu^a(x_\perp) & = & {ig \over 2\pi} \int {d^2z_\perp \over
  (2\pi)^2} {1 \over (x_\perp-z_\perp)^2} {\rm Tr} \left ( T^a
  \Omega^\dagger(x_\perp) \Omega(z_\perp) \right ) \nonumber \\
  \nu^{ab}_{x_\perp,y_\perp} & = & {1 \over \pi} \int {d^2z_\perp
  \over (2\pi)^2} {(x^i-z^i)(y^i-z^i) \over (x_\perp-y_\perp)^2
  (y_\perp-z_\perp)^2} \nonumber \\
  & \times & \left \{ 1+\Omega^\dagger(x_\perp) \Omega(y_\perp)-
  \Omega^\dagger(x_\perp) \Omega(z_\perp)-
  \Omega^\dagger(z_\perp) \Omega(y_\perp) \right \}^{ab}
\end{eqnarray}
As shown in \cite{W02} the coefficients $\nu^{ab}$ and $\nu^a$ are related to each other by the following functional identity
\begin{equation}\label{ward}
{1 \over 2} \int d^2y_\perp \, {\delta \nu^{ab} (x_\perp,y_\perp) \over \delta \alpha^b (y_\perp) } = \nu^a(x_\perp)
\end{equation}
The functional identity (\ref{ward}) allows to rewrite the evolution equation (\ref{RGEa}) in the elegant
Hamiltonian form \cite{W02}
\begin{eqnarray}\label{RGEah}
  {\partial W_{\tau} [\alpha] \over \partial \tau} & = &
  \left \{
  \int {d^2z_\perp \over 2\pi} J^i_a(z_\perp) J^i_a(z_\perp)
  \right \} \,
  W_{\tau}[\alpha] \equiv - H W_{\tau}[\alpha] \nonumber \\
  J^i_a(z_\perp) & = & i \int {d^2x_\perp \over 2\pi} {z^i-x^i \over (z_\perp-x_\perp)^2}
  \left( 1-\Omega^\dagger(z_\perp) \Omega(x_\perp) \right )_{ab}
  {\delta \over \delta \alpha_{\tau}^b(x_\perp)}
 \end{eqnarray}

\begin{itemize}
\item{In the weak field limit (weak $\alpha$) one recovers the BFKL equation \cite{BFKL}.}
\item{Evolving the $n$ - point correlator with the BFKL kernel ${\cal K}_{BFKL}$ leads to the BKP equations.}
\item
 {
 Of special interest is the evolution of ${\cal V}(x_\perp,y_\perp)={\rm Tr} \left( \Omega^\dagger(x_\perp)
 \Omega(y_\perp)\right)$. It is described by the Balitsky-Kovchegov equation \cite{B96,K9900}:
 \begin{equation}\label{evnu}
 {\partial {\cal V}(x_\perp,y_\perp) \over \partial \tau} = -{\alpha_s \over 2 \pi^2} \int d^2 z_\perp
 {(x_\perp-y_\perp)^2 \over (x_\perp-z_\perp)^2 (y_\perp-z_\perp)^2}
 \left( N_c {\cal V}_{xy} - {\cal V}_{xz} {\cal V}_{zy} \right)
 \end{equation}
 }
\item
 {
 Using Eq. (\ref{evnu}) one can write an evolution equation for the scattering amplitude of color dipole on the (gluonic field of the)
 target ${\cal N}(x_\perp,y_\perp)$:
 \begin{eqnarray}
 {\cal N} (x_\perp,y_\perp) & = & {1 \over N_c} {\rm Tr} \left(1- \Omega^\dagger(x_\perp) \Omega(y_\perp) \right) \\
  {\partial {\cal N} (x_\perp,y_\perp) \over \partial \tau} & = &  \int d^2 z_\perp
 {(x_\perp-y_\perp)^2 \over (x_\perp-z_\perp)^2 (y_\perp-z_\perp)^2}
 \left[ {\cal N}_{xz}+{\cal N}_{zy}-{\cal N}_{xy}-{\cal N}_{xz} {\cal N}_{zy} \right] \nonumber
 \end{eqnarray}
 Note that in the Balitsky-Kovchegov equation (\ref{evnu}) the contribution quartic in $\Omega$ is factorized.
 }
\item
 {
        At present the Balitsky - Kovchegov equation (\ref{evnu}) has also been derived in
        \begin{itemize}
        \item{Regge formalism, \cite{BR00}}
        \item{QCD-Regge formalism \cite{BLV04}}
        \end{itemize}
 }

\end{itemize}

\begin{center}
{\bf Some important results}
\end{center}

Let us now mention some important results obtained in the course of developing the color glass condensate physics.

\begin{itemize}
 \item
 {
 Nonlinear effects taken into account in (\ref{evnu}) ensure unitarity restoration for the scattering at fixed
 impact parameter only \cite{KW1,KW2,KW3}. The inelastic cross-section is still characterized by the powerlike growth
 \begin{equation}
 \sigma_{inel} = \pi R^2_{target} + 2 \pi R_{target} x_0 {\rm e }^{\omega \tau}
 \end{equation}
 The origin of the unitarity violation is the powerlike decay of the gluon fields generated by the multipole
 configurations under consideration. To enforce the Froissart bound one has to ensure the exponential decay of the
 gluon fields in the impact parameter plane, which is equivalent to the existence of the mass gap of the theory.
 Obviously, no perturbative framework can describe such an effect.
 }
 \item
 {
  The functional form of the above-described tree-level solution remains unchanged in the course of RG evolution. All changes
  can conveniently be parametrized by evolving the saturation scale $Q^2_s$ \cite{AM99,IM01,IIM02a}:
  \begin{eqnarray}
  Q^2_s(b_\perp,\tau) & = & Q^2_s(b_\perp,\tau_0) {\rm e}^{\lambda \alpha_s (\tau-\tau_0)} \nonumber \\
  {\cal N} (x_\perp,y_\perp) & = & 1-{\rm exp} \left[ -r_\perp^2 Q^2_s(b_\perp,\tau) \right]
  \end{eqnarray}
  Physically this situation can be described as a geometrical scaling of observable cross-sections, when all
  physical quantities depend on energy through $Q^2_s(\cdots,\tau)$ in some range of transverse momenta:
  \begin{equation}
  \sigma=\sigma \left( Q^2/Q^2_s \right), \,\,\,\,\,\,\,\,\,\,\,\,\,\ Q^2 \leq Q^4_s/\Lambda^2_{QCD}
  \end{equation}
 }
 \item{Phenomenological implications of the above - described "$Q^2_s$" - physics are being actively studied for
 ultrarelativistic heavy ion collisions, deep inelastic scattering on nuclei, etc. }

\end{itemize}

\begin{center}
{\bf Conclusions}
\end{center}

Let us conclude with summarizing once again the main points touched in the talk and formulate some major unresolved issues:

\begin{itemize}
\item{Quasiclassical RG formalism provides a compact and transparent view on nonlinear effects in QCD at high energies}
\item
 {
 Nonlinear effects in leading logarithmic approximation provide unitarity restoration at fixed impact parameter only.
 This leaves us with a number of important unsolved questions. Two of them are:
 \begin{itemize}
 \item{How many pomerons does one need to describe high energy QCD?}
 \item{Does perturbative unitarity exist?}
 \end{itemize}
 }
\end{itemize}

The work was supported by RFBR grant 04-02-16880 and the Scientific Schools Support Grant
1976.2003.02.


\begin{thebibliography}{99}%

\bibitem{ILM02}
E.~Iancu, A.~Leonidov, L.~McLerran, {\it The Colour Glass Condensate: An Introduction}, [{\bf
arXiv:hep-ph/0202270}]

\bibitem{IV03}
E.~Iancu, R. Venugopalan, {\it The Color Glass Condensate and High Energy Scattering in QCD}, [{\bf
arXiv:hep-ph/0303204}]

\bibitem{JKLW97}
J.~Jalilian-Marian, A.~Kovner, A.~Leonidov, H.~Weigert, {\it Nucl. Phys.}\, {\bf B504} (1997), 415.


\bibitem{JKLW99a}
J.~Jalilian-Marian, A.~Kovner, A.~Leonidov, H.~Weigert, {\it Phys. Rev.}\,{\bf D59} (1999), 014014.

\bibitem{ILM00}
E.~Iancu, A.~Leonidov, L.~McLerran, {\it Nucl. Phys.}\,{\bf A692}(2001), 583.

\bibitem{ILM01}
E.~Iancu, A.~Leonidov, L.~McLerran, {\it Phys. Lett.}\,{\bf B510} (2001), 45.

\bibitem{FILM02}
E.~Fereiro, E.~Iancu, A.~Leonidov, L.~McLerran, {\it Nucl. Phys.}\,{\bf A703} (2002), 489.

\bibitem{BFKL}
L.N.~Lipatov, {\it Sov. Journ. Nucl. Phys.}\,{\bf 23} (1976),
338;\\
E.A.~Kuraev, L.N.~Lipatov, V.S.~Fadin, {\it Sov. Phys. JETP}\,
{\bf 45} (1977), 199;\\
Ya.Ya.~Balitsky, L.N.~Lipatov, {\it Sov. Journ. Nucl. Phys.}\, {\bf 28} (1978), 822.

\bibitem{Li95}
L. N. Lipatov, {\it Nucl. Phys}\,{\bf B452} (1995), 369.

\bibitem{Li97}
L. N. Lipatov, {\it Phys.\ Rept.}\,{\bf 286} (1997), 131.

\bibitem{B96}
I.~Balitsky, {Nucl. Phys.}\,{\bf B463}(1996), 99.

\bibitem{MV94}
L.~McLerran, R.~Venugopalan, {\it Phys. Rev.}\, {\bf D49} (1994),
3352;\\
{\it ibid.} {\bf D50} (1994), 2225.

\bibitem{JKMW97}
J. Jalilian-Marian, A. Kovner, L. McLerran, H. Weigert, {\it Phys.Rev.}\,{\bf D55} (1997), 5414.

\bibitem{KM98}
Yu.V.~Kovchegov, A.H.~Mueller, {\it Nucl. Phys.}\,{\bf B259} (1998), 451.

\bibitem{W02}
H.~Weigert, {\it Nucl.\ Phys.}\ {\bf A703} (2002), 823

\bibitem{K9900}
Yu. Kovchegov, {\it Phys. Rev.}\,{\bf D60}(1999), 034008;\\
{\it ibid.} \,{\bf D61}(2000), 074018.

\bibitem{BR00}
M.~Braun, {\it Eur. Phys. Journ.}\, {\bf C16} (2000), 337.

\bibitem{BLV04}
J.~Bartels, L.N.~Lipatov, G.P.~Vacca, "Interactions of Reggeized Gluons in the Moebius Representation", {\bf [
arXiv:hep-ph/0404110.]}

\bibitem{KW1}
A.~Kovner, U.~Wiedemann, {\it Phys.\ Rev.}\ {\bf D66} (2002), 051502

\bibitem{KW2}
A.~Kovner, U.~Wiedemann, {\it Phys.\ Rev.}\ {\bf D66} (2002), 034031

\bibitem{KW3}
A.~Kovner, U.~Wiedemann, {\it Phys.\ Lett.}\ {\bf B551} (2003), 511

\bibitem{AM99}
A.H.~Mueller, {\it Nucl.\ Phys.}\ {\bf B558} (1999), 285.

\bibitem{IM01}
E.~Iancu, L.~McLerran, {\it Phys.\ Lett.}\ {\bf B510} (2001), 145.

\bibitem{IIM02a}
E.~Iancu, K.~Itakura, L.~McLerran, {\it Nucl.\ Phys.}\ {\bf A708} (2002), 327.

\end{thebibliography}
\end{document}